\documentclass[a4paper,12pt]{article}
\usepackage[margin=2.5cm]{geometry}
\usepackage{amsmath}
\usepackage{graphicx}
\usepackage{bm}
\usepackage{amssymb}
\usepackage{hyperref}
\usepackage[style=apa,backend=biber,natbib=true]{biblatex}
\title{Stochastic Backscatter Model for Unstructured Solvers}
\author{Angelo Passariello}

\begin{document}
\maketitle

\tableofcontents

\section{Problem statement}
The Stochastic BackScatter (SBS) model involves the generation of a set of random variables featured by specific correlations in space and time \parencite{kok2017stochastic}. To obtain these variables a three-step procedure is employed:
\begin{enumerate}
	\item  A three-dimensional field (say, $dV_i$) of completely uncorrelated normally-distributed stochastic quantities is generated first.
	\item An implicit smoothing of $dV_i$ is performed to add an exponentially-decaying spatial correlation. The spatially-correlated field will be denoted as $dW_i$. The scale for decay is proportional to the LES filter width.
	\item Three Langevin equations are solved to add an exponentially-decaying temporal correlation to the smoothed field $dW_i$. The latter appears as source term in the Langevin equations. 
\end{enumerate}
The implicit smoothing (point 2 of the above procedure) is carried out by solving three nested elliptic problems in the computational space ($x_{\scalebox{0.6}{I}}x_{\scalebox{0.6}{J}}x_{\scalebox{0.6}{K}}$), as shown below:
\begin{equation}
	d\bm x\left(1-b^2\frac{\partial}{\partial x_{\scalebox{0.6}{K}}^2}\right)\left(1-b^2\frac{\partial}{\partial x_{\scalebox{0.6}{J}}^2}\right)\left(1-b^2\frac{\partial}{\partial x_{\scalebox{0.6}{I}}^2}\right)dW_i = 8b^{3/2}dV_i
	\label{eq:smoothing1}
\end{equation}
In the previous equation, $b$ represents a scalar (which is proportional to the LES filter width) and $d\bm x$ is an elemental volume. It is clear that this approach is functional for structured grids only. To extend the SBS model to unstructured CFD solvers, a Laplacian smoothing could be employed instead:
\begin{equation}
	d\bm x\left(1-b^2\nabla^2\right)\,dW_i = 8b^{3/2}dV_i
	\label{eq:smoothing2}
\end{equation}
In fact, Eq.~\eqref{eq:smoothing1} is an approximation of Eq.~\eqref{eq:smoothing2} with an error of order $\mathcal{O}(b^4)$.
However, the spatial correlation resulting from Eq.~\eqref{eq:smoothing2} must be derived mathematically. This is the objective of the present dissertation.

\section{Green's function for the Laplacian smoothing}
Let $G(\bm{x})$ be the three-dimensional Green's function satisfying the following equation:
\begin{equation}
	G(\bm{x})-b^2\,\nabla^2 G(\bm{x}) = \delta(\bm{x})
	\label{eq:smoothing}
\end{equation}
\noindent
where $\delta(\bm{x})$ denotes the three-dimensional Dirac delta function and $b$ is a scalar factor. 
Let $\Hat{G}(\bm{\kappa})$ indicate the Fourier transform of $G$, defined as:
\begin{equation}
	\Hat{G}(\bm\kappa)=\int_{\mathbb{R}^3} e^{-i\,\bm\kappa\cdot\bm x}\,G(\bm x) \,d \bm x
\end{equation}
The following convention is employed for the inverse Fourier transform:
\begin{equation}
	G(\bm x) = \frac{1}{(2\pi)^3}\int_{\mathbb{R}^3} e^{i\,\bm\kappa\cdot\bm x}\,\Hat{G}(\bm \kappa) \,d \bm \kappa
\end{equation}
Here, $d\bm x$ and $d\bm \kappa$ represent two infinitesimal ``volumes'' in the spatial and frequency domain respectively.
Fourier-transforming both sides of Eq.~\eqref{eq:smoothing}, the following expression is obtained:
\begin{equation}
	(1+b^2|\bm\kappa|^2)\,\Hat{G}(\bm\kappa) =1
\end{equation}
\noindent
from which $\Hat{G}$ can easily be derived:
\begin{equation}
	\Hat{G}(\bm\kappa)=\frac{1}{1+b^2|\bm\kappa|^2}
\end{equation}
\noindent
Transforming $\Hat{G}$ back to the spatial domain, one gets:
\begin{equation}
	G(\bm x)=\frac{1}{(2\pi)^3}\int_{\mathbb{R}^3}\frac{e^{i\bm\kappa\cdot\bm x}}{1+b^2|\bm\kappa|^2}\,d\bm\kappa
	\label{eq:green}
\end{equation}
\noindent
The computation of this integral is not immediate. First of all, let us rotate the wavenumber vector (i.e., $\bm\kappa$) using a generic matrix $\mathbf{R}$ to obtain a new wavenumber vector, $\bm\kappa'$: 
\begin{equation}
	\bm\kappa'=\mathbf{R}\bm\kappa
\end{equation}
Using a well-known property of rotation matrices (i.e., $\mathbf{R}^{-1}=\mathbf{R}^T$), $\bm \kappa$ can be expressed as a function of $\bm\kappa'$:
\begin{equation}
	\bm\kappa=\mathbf{R}^T\bm\kappa'
	\label{eq:krot}
\end{equation}
The modulus of $\bm\kappa$ is obviously preserved by the rotation. This can be shown as follows:
\begin{equation}
	\bm\kappa^T\bm\kappa=\bm\kappa^T\mathbf{R}^T\bm\kappa'
\end{equation}
\begin{equation}
	\bm\kappa^T\bm\kappa=(\mathbf{R}\bm\kappa)^T\bm\kappa'
\end{equation}
\begin{equation}
	\bm\kappa^T\bm\kappa=(\bm\kappa')^T\bm\kappa'
\end{equation}
\begin{equation}
	|\bm\kappa|^2=|\bm\kappa'|^2
\end{equation}
\noindent
In a similar way, it can be proved that: $d\bm k = d\bm\kappa'$.
The exponential in Eq.~\eqref{eq:green} deserves more attention.
By substituting Eq.~\eqref{eq:krot} in Eq.~\eqref{eq:green}, the aforementioned term becomes:
\begin{equation}
	e^{i\,(\mathbf{R}^T\bm\kappa')\cdot \bm x}=e^{i\,\bm\kappa'\cdot(\mathbf{R}\bm x)}
	\label{eq:dotprod}
\end{equation}
\noindent
In conclusion:
\begin{equation}
	G(\bm x)=\frac{1}{(2\pi)^3}\int_{\mathbb{R}^3}\frac{e^{i\bm\kappa'\cdot(\mathbf{R}\bm x)}}{1+b^2|\bm\kappa'|^2}\,d\bm\kappa'
	\label{eq:green3}
\end{equation}
We have proved that $G(\bm x)$ is invariant to rotations.
Consequently, the matrix $\mathbf R$ can be constructed in such a way to simplify the computation of the integral in Eq.~\eqref{eq:green3}. 
Let us choose $\mathbf R$ such that the vector $\bm x$ becomes aligned with the $z$-axis of the physical space:
\begin{equation}
	\mathbf{R}\bm x = |\bm x|\,\hat{\bm z}
\end{equation}
\noindent
Then, if a spherical coordinate system is employed, the dot product in Eq.~ \eqref{eq:dotprod} can be expressed as:
\begin{equation}
	e^{i\,\bm\kappa'\cdot(|\bm x|\,\hat{\bm z})} = e^{i\,|\bm\kappa|\,|\bm x|\,\cos\theta}
	\label{eq:exp}
\end{equation}
where $\theta$ denotes the polar angle (i.e., $\theta=\cos^{-1}(\bm\kappa'\cdot\bm{\hat z}/|\bm \kappa'|)$). 
The volume element $d\bm\kappa$ in spherical coordinates is written as:
\begin{equation}
	d\bm k= |\bm\kappa|^2\,\sin\theta\,d\theta\,d\phi\,d|\bm\kappa|
	\label{eq:volel}
\end{equation} 
where $\phi$ represents the azimuthal angle.
By substituting Eqs.~\eqref{eq:exp}-\eqref{eq:volel} in Eq.~\eqref{eq:green3}, one finally obtains:
\begin{equation}
	G(\bm x) = \frac{1}{(2\pi)^3}\int_0^{2\pi}d\phi \,\int_0^{+\infty}\int_0^\pi \frac{e^{i\,|\bm\kappa|\,|\bm x|\,\cos\theta}}{1+b^2|\bm\kappa|^2}|\bm\kappa|^2\sin\theta\,d\theta\,d|\bm\kappa|
	\label{eq:green2}
\end{equation}
Let us solve the angular part first:
\begin{equation}
	\int_0^{2\pi}d\phi = 2\pi\;
	\label{eq:tmp1}
\end{equation}
\begin{equation}
	\int_0^\pi e^{i\,|\bm\kappa|\,|\bm x|\,\cos\theta}\sin\theta\,d\theta\xrightarrow{u=\cos\theta}
	\int_{-1}^1 e^{i\,|\bm\kappa|\,|\bm x|\,u}\,du = \left[\frac{e^{i\,|\bm\kappa|\,|\bm x|\,u}}{i\,|\bm\kappa|\,|\bm x|}\right]_{-1}^{1}= \frac{2\,\sin(|\bm\kappa|\,|\bm x|)}{|\bm\kappa|\,|\bm x|}
	\label{eq:tmp2}
\end{equation}
By substituting Eqs.~\eqref{eq:tmp1}-\eqref{eq:tmp2} in Eq.~\eqref{eq:green2}:
\begin{equation}
\begin{split}
	G(\bm x)&=\frac{1}{(2\pi)^3}\cdot(2\pi)\cdot\int_0^{+\infty} \frac{|\bm\kappa|^2}{1+b^2|\bm\kappa|^2}\,\frac{2\sin(|\bm\kappa||\bm x|)}{|\bm\kappa| |\bm x|}\,d|\bm\kappa| \\
	&=\frac{1}{2\pi^2|\bm x|}\int_0^{+\infty}\frac{|\bm\kappa|\,\sin(|\bm\kappa||\bm x|)}{1+b^2|\bm\kappa|^2}\,d|\bm\kappa|\\
	&=\frac{1}{2\pi^2b^2|\bm x|}\int_0^{+\infty}\frac{(b|\bm\kappa|)\,\sin\left(b|\bm\kappa|\frac{|\bm x|}{b}\right)}{1+b^2|\bm\kappa|^2}\,d(b|\bm\kappa|)
\end{split}
\end{equation}
Using the residue theorem (see the Appendix for details), it can be proved that:
\begin{equation}
	\int_0^{+\infty}\frac{u\,\sin(\alpha u)}{1+u^2}\,du=\frac{\pi}{2}\,e^{-\alpha}
\end{equation}
Here, in particular: $u=b|\bm\kappa|$, $\alpha=|\bm x|/b$.
The expression of the Green's function can finally be computed:
\begin{equation}
	G(\bm x)=\frac{1}{2\pi^2b^2|\bm x|}\cdot\frac{\pi}{2}\,e^{-|\bm x|/b}=\frac{e^{-|\bm x|/b}}{4\pi b^2|\bm x|}
\end{equation}

\section{Spatial correlation of the stochastic differentials}
Known $G(\bm x)$, the general solution for the implicit Laplacian smoothing in Eq.~\eqref{eq:smoothing2} can be written as:
\begin{equation}
	dW_i(\bm x,t)=\int_{\mathbb{R}^3} G(\bm x - \bm x')\,8b^{3/2}dV_i(\bm x',t)
\end{equation}
If $dV_i$ is completely uncorrelated both in space and in time, then it follows that:
\begin{equation}
\begin{split}
	\langle dW_i(\bm x, t)\,dW_j(\bm y,s) \rangle &= 64\,b^3 \int_{\mathbb{R}^3}\int_{\mathbb{R}^3}G(\bm x - \bm x')G(\bm y - \bm y')\langle dV_i(\bm x', t)\,dV_j(\bm y',s) \rangle \\
	&= 64\,b^3\delta_{ij}\delta(t-s)\,\int_{\mathbb{R}^3} G(\bm x - \bm x')G(\bm y - \bm x')\,d\bm x'\\
\end{split}
\label{eq:corr}
\end{equation}
where the angular brackets denote ensemble averaging.
It is trivial to show that the integral in Eq.~\eqref{eq:corr} is invariant to translations of $\bm x$ and $\bm y$. Indeed, letting $\bm a$ be a generic vector, one can write:
\begin{equation}
\begin{split}
	\langle dW_i(\bm x + \bm a, t)\,dW_j(\bm y + \bm a,s) \rangle &\propto \int_{\mathbb{R}^3} G(\bm x - \bm x')G(\bm y - \bm x')\,d\bm x'\\
	&\propto\int_{\mathbb{R}^3} G(\bm x + \bm a - \bm x')G(\bm y + \bm a - \bm x')\,d\bm x'\\
	&\propto\int_{\mathbb{R}^3} G(\bm x - (\bm x'-\bm a))G(\bm y - (\bm x' - \bm a))\,d(\bm x'- \bm a)
\end{split}
\end{equation}
Without loss of generality, one can thus translate the coordinate system of $-\bm y$. The following integral must thus be solved:
\begin{equation}
	\int_{\mathbb{R}^3} G(\bm x - \bm y - \bm x')G(\bm x')\,d\bm x'
\end{equation}
By definition, the above equation represents the convolution of the Green function $G(\bm x - \bm y)$ with itself. For notational simplicity, the following substitutions are made: $\bm x - \bm y \leftrightarrow \bm r$, $\bm x' \leftrightarrow \bm r'$. 
It is known that the Fourier transform of the convolution product of two functions is equal to the product of their respective Fourier transforms. Namely:
\begin{equation}
	\mathcal{F}\left\{\int_{\mathbb{R}^3} G(\bm r - \bm r')G(\bm r')\,d\bm r'\right\} = [\Hat{G}(\bm\kappa)]^2
\end{equation}
Therefore, we can compute the convolution integral by anti-transforming $[\Hat{G}(\bm\kappa)]^2$ (whose expression is known from the previous Section). To be precise, we have to calculate:
\begin{equation}
	\frac{1}{(2\pi)^3}\int_{\mathbb{R}^3}\frac{e^{i\bm\kappa\cdot\bm r}}{(1+b^2|\bm\kappa|^2)^2}\,d\bm\kappa
\end{equation}
By following the same approach used to compute the Green's function, it can be shown that:
\begin{equation}
	\frac{1}{(2\pi)^3}\int_{\mathbb{R}^3}\frac{e^{i\bm\kappa\cdot\bm r}}{(1+b^2|\bm\kappa|^2)^2}\,d\bm\kappa=\frac{e^{-|\bm r|/b}}{8\pi b^3}
\end{equation}
To conclude, the following spatial correlation of the stochastic differentials is obtained:
\begin{equation}
	\langle dW_i(\bm x, t)\,dW_j(\bm y,s) \rangle = 64\,b^3\delta_{ij}\delta(t-s) \cdot \frac{e^{-|\bm x - \bm y|/b}}{8\pi b^3} = \frac{8}{\pi}\delta_{ij}\delta(t-s)e^{-|\bm x - \bm y|/b}
\end{equation}
For the sake of completeness, the spatial correlation obtained from Eq.~\eqref{eq:smoothing1} is also shown \parencite{kok2017stochastic}:
\begin{equation}
	\langle dW_i(\bm x, t)\,dW_j(\bm y,s) \rangle = \delta_{ij}\delta(t-s)\left(e^{-|\bm x - \bm y|^2/(2b^2)}+\mathcal{O}(|\bm x-\bm y|^3)\right)
\end{equation}
It must be emphasised that the exponential decay in space that is achieved with the implicit Laplacian smoothing is comparatively weak.

\addcontentsline{toc}{section}{Appendix: application of the residue theorem}
\section*{Appendix: application of the residue theorem}
It must be proved that:
\begin{equation}
	\int_0^{+\infty}\frac{u\,\sin(\alpha u)}{1+u^2}\,du=\frac{\pi}{2}\,e^{-\alpha}
	\label{eq:residueint}
\end{equation}
To this aim, the following remarkable results from Complex Analysis will be used:\\[2pt]
\hrule
\vspace{2pt}
\hrule
\vspace{10pt}
\noindent
\textbf{Residue theorem.} Let $\mathcal{U}$ be a simply connected open subset of the complex plane containing a finite list of points $a_1,...,a_n$ and $f$ be a function holomorphic on $\mathcal{U}_0=\mathcal{U}\backslash\{a_1,...,a_n\}$.
Letting $\gamma$ be a closed rectifiable curve in $\mathcal{U}_0$, and denoting the residue of $f$ at each point $a_k$ by $\mathrm{Res}(f,a_k)$, and the winding number of 
$\gamma$ around $a_k$ by $I(\gamma,a_k)$, the line integral of $f$ around $\gamma$ is equal to $2\pi i$ times the sum of residues, each counted as many times as $\gamma$ winds around the respective point:
\begin{equation}
	\oint_\gamma f(z)\,dz = 2\pi i\sum_{k=1}^n I(\gamma,a_k)\,\mathrm{Res}(f,a_k)
\end{equation}
If $\gamma$ is a positively oriented simple closed curve, $I(\gamma,a_k)$ is 
1 if $a_k$ is in the interior of $\gamma$ and 0 if not, therefore:
\begin{equation}
	\oint_\gamma f(z)\,dz = 2\pi i\sum\mathrm{Res}(f,a_k)
	\label{eq:int2}
\end{equation}
with the sum over those $a_k$ inside $\gamma$.\\[2pt]\hrule
\vspace{2pt}
\hrule
\vspace{10pt}
\noindent
\textbf{Jordan's lemma.} 
Consider a complex-valued, continuous function $f$, defined on a semicircular contour
$C_R=\{R\,e^{i\theta }\mid \theta \in [0,\pi ]\}$ of positive radius $R$ lying in the upper half-plane, centered at the origin. If the function $f$ is of the form $f(z)=e^{i\alpha z}g(z),\quad z\in C$ with a positive parameter $\alpha$, then Jordan's Lemma states the following upper bound for the contour integral:
\begin{equation}
	\left|\int _{C_{R}}f(z)\,dz\right|\leq {\frac {\pi }{\alpha}}M_{R}\quad {\text{where}}\quad M_{R}:=\max _{\theta \in [0,\pi ]}\left|g\left(R\,e^{i\theta }\right)\right|.
\end{equation}
with equality when $g$ vanishes everywhere, in which case both sides are identically zero. An analogous statement for a semicircular contour in the lower half-plane holds when $\alpha<0$.
\\[2pt]\hrule
\vspace{2pt}
\hrule
\vspace{10pt}
\noindent
To prove the identity in Eq.~\eqref{eq:residueint}, let us first consider the following integral:
\begin{equation}
	C(\alpha)=\int_{-\infty}^{+\infty}\frac{e^{i\alpha u}}{1+u^2}\,du
	\label{eq:int1}
\end{equation}
with $\alpha$ positive and $u\in\mathbb{R}$. We can evaluate $C(\alpha)$ through the residue theorem. We have to define a convenient curve $\gamma$ in the complex plane. The latter can be constructed as the union of the semi-circumference of radius $R$ centred in the origin of the complex plane and of the segment $[-R,R]$ lying on the real axis. 
Referring to the statement of Jordan's lemma and looking at the integrand in Eq.~\eqref{eq:int1}, we notice that:
\begin{equation}
	f(z)=e^{i\alpha z}\,g(z)=\frac{e^{i\alpha z}}{1+z^2}\longrightarrow
	g(z)=\frac{1}{1+z^2}
\end{equation}
Computing the maximum of the modulus of $g$ is trivial: 
\begin{equation}
	M_R=\frac{1}{1+R^2}
\end{equation}
According to Jordan's lemma:
\begin{equation}
	0 \leq \left|\int_{C_R} \frac{e^{i\alpha z}}{1+z^2}\,dz\right| \leq \frac{1}{1+R^2}
\end{equation}
As a consequence:
\begin{equation}
	\lim_{R\to +\infty}\left|\int_{C_R} \frac{e^{i\alpha z}}{1+z^2}\,dz\right| =0
\end{equation}
 This means that the application of the residue theorem on $\gamma$ (when the radius $R$ tends to infinity) will directly yield $C(\alpha)$. The only pole of $f$ included in $\gamma$ is $z=i$. We can easily evaluate the residue in that point according to the definition:
 \begin{equation}
 	\mathrm{Res}(f,i)=\lim_{z\to i}\,(z-i) f(z) = \lim_{z\to i} \frac{e^{i\alpha z}}{z+i} = \frac{e^{-\alpha}}{2i}
 \end{equation}
 Recalling Eq.~\eqref{eq:int2}, one can write:
 \begin{equation}
 	\int_{-\infty}^{+\infty} \frac{e^{i\alpha u}}{1+u^2}\,du = 2\pi i \, \frac{e^{-\alpha}}{2i} = \pi e^{-\alpha}
 \end{equation}
 Since the integral is real, only the real part of the integrand function contributes non-zero values, namely: 
 \begin{equation}
 	\int_{-\infty}^{+\infty}\frac{e^{i\alpha u}}{1+u^2}\,du =\int_{-\infty}^{+\infty}\frac{\cos(\alpha u)}{1+u^2}\,du = \pi e^{-\alpha}
 \end{equation}
 Furthermore, the real part of the integrand function is even. As a consequence:
 \begin{equation}
 	\int_0^{+\infty} \frac{\cos(\alpha u)}{1+u^2}\,du =\frac{1}{2}\int_{-\infty}^{+\infty} \frac{\cos(\alpha u)}{1+u^2}\,du= \frac{\pi}{2}e^{-\alpha}
 \end{equation}
 By deriving both sides of the previous equation with respect to $\alpha$, one obtains:
 \begin{equation}
 	\int_0^{+\infty}\frac{u \sin(\alpha u)}{1+u^2}\,du=\frac{\pi}{2}e^{-\alpha}
 \end{equation}
 which is exactly the identity we wanted to prove.

\printbibliography[heading=bibintoc,title={References}]

\end{document}